# Complexity and white-dwarf structure

J. Sañudo[1,*], and A.F. Pacheco[2]

[1] *Departamento de Física, Universidad de Extremadura, 06071 Badajoz, Spain.*

[2] *Facultad de Ciencias and BIFI, Universidad de Zaragoza, 50009 Zaragoza, Spain.*

## ABSTRACT

From the low-mass non-relativistic case to the relativistic limit, the density profile of a white dwarf is used to evaluate the $C_{CLM}$ complexity measure [1]. Similarly to the recently reported atomic case where, by averaging shell effects, complexity grows with the atomic number [2-4], here complexity grows as a function of the star mass reaching a maximum finite value in the Chandrasekhar limit.

PACS numbers: 71.10.Ca, 89.70. - a

___________________________________________

*\* Corresponding author*.

E-mail addresses: jsr@unex.es (J. Sañudo), amalio@unizar.es (A.F. Pacheco).

Recent years have witnessed a great deal of work in the field of complexity and information theories because of their likely relevance to the understanding of biological processes at their basic level [5]. However, in spite of all this effort, a satisfactory universal definition of complexity is still lacking and different indicators emphasize different points of view or perspectives on this issue. Although, as said above, the main motivation of these developments was their application to biological systems, there have recently been a number of studies of complexity in atomic and molecular systems. When applied to these systems [2-4, 6-8] the statistical complexity measures [1, 9] have the form of the habitual matrix density functional. In these systems, a bigger or smaller value of the complexity can be a manifestation of shell structure, correlations, importance of relativistic effects, etc. Motivated by the application of these ideas to heavy atoms, here we apply them to a real self-gravitating system, a white dwarf. As far as we know, this is the first study of complexity in an astrophysical object.

Among all the statistical complexity measures, we will use that proposed by López-Ruiz, Calbet and Mancini [1] which is defined as

$$C_{LMC} = H \cdot D \quad , \tag{1}$$

where $H$ represents the information content of the system [10]

$$H = \frac{1}{2\pi e} e^{2S/3} \quad . \tag{2}$$

The magnitude $S$, given by

$$S = -\int \hat{\rho}(\vec{r}) \log\left(b_0^3 \hat{\rho}(\vec{r})\right) d\vec{r} \quad , \tag{3}$$

is the Shannon information entropy [11]. And $D$, given by

$$D = b_0^3 \int \hat{\rho}^2(\vec{r}) d\vec{r} \quad , \tag{4}$$

is called the disequilibrium [1]. The length scale $b_0$ depends on the mass of the star and will be defined later.

From their definitions, $H$ measures the information and $D$ the disequilibrium or distance to the most probable state. In a perfect ordered system, $H$ is null, and in a state of maximum disorder, $D$ is null. Thus, in a sense, the measure $C_{LMC}$ simultaneously probes the distance of a system to the state of perfect order and to that of perfect disorder. $C_{LMC}$ has certainly received criticism [12] but we will use it because of its simplicity and intuitiveness.

Now we will recall the basic equations of a white dwarf [13]. Henceforth, we will use the following list of definitions and numerical values.

$\hbar$ : reduced Planck constant.

$c$ : light velocity.

$G = 6.707 \times 10^{-39} \, \hbar c \left( \dfrac{Gev}{c^2} \right)^{-2}$, gravitational constant.

$m_e = 0.511$ MeV/$c^2$, mass of the electron.

$m = 939$ $MeV/c^2$, mass of the nucleon

$M$, mass of the star.

$M_\odot = 1.98892 \times 10^{30}$ $Kg$, solar mass.

$R$, radius of the star.

$r$, distance to the centre of the star.

ϕ, gravitational field.

*A*, number of electrons ( or protons) in the star.

*f*, number of nucleons per electron. We will take $f = 2$.

$p_F$, Fermi momentum of the electrons.

$n_e$, number of electrons per unit volume.

ρ, mass density in the star, $\rho = (fm + m_e) n_e \approx fm\, n_e$.

Natural units $\hbar = c = 1$, and spherical symmetry will be assumed. The structure of a white dwarf is expressed by three equations and we will use [14] a notation reminiscent of the Thomas-Fermi (TF) model, well known in Atomic Physics [15]. We have first the Poisson equation describing the Newtonian gravitational field.

$$\vec{\nabla}^2 \phi = 4\pi G \rho. \tag{5}$$

Second, we have the equation of hydrostatic equilibrium,

$$\sqrt{p_F^2 + m_e^2} - m_e + fm\phi = -C, \tag{6}$$

where *C* is a constant. Equation (6) says that the maximum energy that an electron and *f* nucleons can have is independent of *r*. This equation is ordinarily expressed as a balance between the gravitational force, mostly generated by the nucleons, acting towards the centre of the star and the Fermi pressure gradient, generated by electrons, acting outwards. Finally, we have the equations of state [15]

$$n_e = \frac{p_F^3}{3\pi^2}, \tag{7}$$

that relates the electron Fermi momentum with the electron density. Eliminating $p_F$ and $n_e$ and defining the dimensionless variables χ and *x* in the form

$$b = \left(\frac{9\pi^2}{128}\right)^{1/3} \frac{1}{GM^{1/3}(fm)^{5/3} m_e} \approx \left(\frac{9\pi^2}{128}\right)^{1/3} \frac{A^{-1/3}}{f^2} b_0, \quad b_0 = \frac{1}{Gm^2 m_e}, \tag{8}$$

we obtain the following differential equation [14]

$$\ddot{\chi}(x) = -\frac{\chi^{3/2}}{x^{1/2}}\left\{1 + \theta \cdot \frac{\chi}{x}\right\}^{3/2}, \qquad (9)$$

where

$$\theta = \frac{GMfm}{2m_e b} \qquad (10)$$

The physical solution of Eq.(9) has to verify two conditions. The fact that the gravitational field at the centre of the star is not infinite imposes

$$\chi(0) = 0, \qquad (11)$$

and the fact that the integral of the density profile has to be equal to $M$ imposes

$$\begin{aligned} x_M \cdot |\dot{\chi}(x_M)| &= 1, \\ x_M &\equiv R/b \end{aligned} \qquad (12)$$

In terms of the solution, $\chi$, the mass density is given by

$$\begin{aligned} \rho(x) &= \frac{M}{4\pi b^3}\frac{\chi^{3/2}}{x^{3/2}}\left\{1 + \theta \cdot \frac{\chi}{x}\right\}^{3/2}, \\ \hat{\rho}(x) &\equiv \rho(x)/M \end{aligned} \qquad (13)$$

and the limit of Chandrasekhar [13,14] imposes that only for values

$$\theta^{3/2} \leq \theta_{lim}^{3/2} = 4.0729 \qquad (14)$$

do there exist physical solutions.

Using the numerical values quoted at the beginning, the value of θ is

$$\theta = 0.24087\, f^{8/3}\, (M/M_\odot)^{4/3} \qquad (15)$$

The results for *S, H, D* and $C_{LMC}$ have been plotted in Fig.1, *a, b, c,* and *d* respectively as a function of the star mass. Their behaviour can be understood in the following way. In the extreme relativistic limit, Eq. (9) reduces to

$$\ddot{\chi}_{lim}(x) = -\theta_{lim}^{3/2} \frac{\chi^3}{x^2} \quad , \tag{16}$$

which is invariant under the scale transformation

$$x \rightarrow y = \lambda x \quad . \tag{17}$$

Which means that its solution lacks a definite radius [13]. Note that on its way to this extreme limit, the radius of a white dwarf decreases as its mass grows. Thus, expressing the limiting behaviour of the relevant magnitudes of our problem in terms of the parameter λ tending to 0, after a light effort we find

$$lim\, S(M \rightarrow M_{lim}) \propto log\lambda^3, \tag{18}$$

$$lim\, H(M \rightarrow M_{lim}) \propto \lambda^3, \tag{19}$$

$$lim\, D(M \rightarrow M_{lim}) \propto \lambda^{-3}, \tag{20}$$

and finally

$$lim\, C_{LMC}(M \rightarrow M_{lim}) = constant \quad . \tag{21}$$

These behaviours are consistent with the graphs shown in Fig.1*a-d* (continuous line). Intuitively speaking, the decrease of *S* and *H* when the mass of the star grows (λ→0) was expected because the more concentrated the star in the space, the smaller the information entropy *S* has to be. And the increasing concentration of the density necessarily causes the growth of *D* because it is nothing more than the density expectation value. Finally, in the plot for the complexity measure, $C_{LMC}$ (Fig.1*d*), one sees its monotonous increase as the star mass grows, ending in the extreme relativistic limit as a finite constant around 1.83. For comparison, it is also shown in Fig.1*a-d* the non-relativistic calculation (dashed line). It should be noted that in the non-relativistic calculation (θ = 0), there is no limit in the mass of the star.

The growth of $C_{LMC}$ with $M$ is qualitatively analogous to what has been reported in atomic physics [2-4] where $C_{LMC}$ grows with the atomic number if one averages the shell structure underlying the periodic table. This qualitative coincidence was also expected because of the strong similarity between both systems. In the two cases one is dealing with degenerate electrons bound by the external potential of the nucleus in the case of the atom, and by their self-gravitating forces in the case of the star. Besides, the forces in both systems are proportional to the inverse of the square of the distance. And finally, the kinematics of the electrons is initially non-relativistic in both systems (low $Z$ and low $M$) becoming more and more relativistic with the increase in the number of particles.

## Acknowledgements

This work was supported in part by the Spanish DGICYT under Project FIS 2005-06237 of the Ministerio Español de Ciencia y Tecnología.

# REFERENCES

[1] R. López-Ruiz, H.L. Mancini, and X. Calbet, *Phys. Lett*. A **209** (1995) 321.

[2] C.P. Panos, K.Ch. Chatzisavvas, Ch.C. Moustakidis, and E.G. Kyrkow, *Phys. Lett*. A **363** (2007) 78.

[3] A. Borgoo, F. De Proft, P. Geerlings, K.D. Sen, Chem. Phys. Lett. **444** (2007) 186.

[4] J. Sañudo and R. López-Ruiz, Int. Rev. of Phys. (2008), *in press*.

[5] R.A. Gatenby and B.R. Frieden, Bull. Math Biol. 69 (2007) 635.

[6] S.E. Massen and C.P. Manos, Phys. Lett. A **246** (1998) 530.

[7] H.E. Montgomery Jr and K.D. Sen, Phys. Lett A **372** (2008) 2271.

[8] J. Sañudo and R. López-Ruiz, J. of Phys. A: Math. Theor. **41** (2008) 265303.

[9] J.S. Shiner, M. Davison, and P.T. Landsberg, Phys. Rev. E **59** (1999) 1459.

[10] A. Dembo, T.A. Cover, and J.A. Thomas, *IEEE Trans. Inf. Theory* **37** (1991) 1501.

[11] C.E. Shannon, *Bell Sys. Tech. J*. **27** (1948) 379; *ibid*. **27** (1948) 623.

[12] D.P. Feldman and J.P. Crutchfield, Phys. Lett A **238** (1998) 244.

[13] See, for example, S.L. Shapiro and S.A. Teukolsky, *Black Holes, White Dwarfs and Neutron Stars* (Wiley, New York, 1983).

[14] M. Membrado and A.F. Pacheco, Astrophys. J. **331** (1988) 394.

[15] See for example. S. Eliezer, A. Ghatak and H. Hora, *An Introduction to Equations of State* (Cambridge University Press, Cambridge, 1986), and references therein.

# CAPTIONS FOR THE FIGURES

Fig.1$a$ - The Shannon information entropy versus the mass of the star. The continuous line represents the relativistic calculation. The dashed line corresponds to the non-relativistic calculation. For details see the text.

Fig.1$b$ - The information content, $H$, versus the mass of the star. The comments in Fig.1$a$ are also valid here.

Fig.1$c$ - The disequilibrium, $D$, versus the mass of the star. The comments in Fig.1$a$ are also valid here.

Fig.1$d$ - The complexity, $C_{LMC}$, versus the mass of the star. The comments in Fig.1$a$ are also valid here.

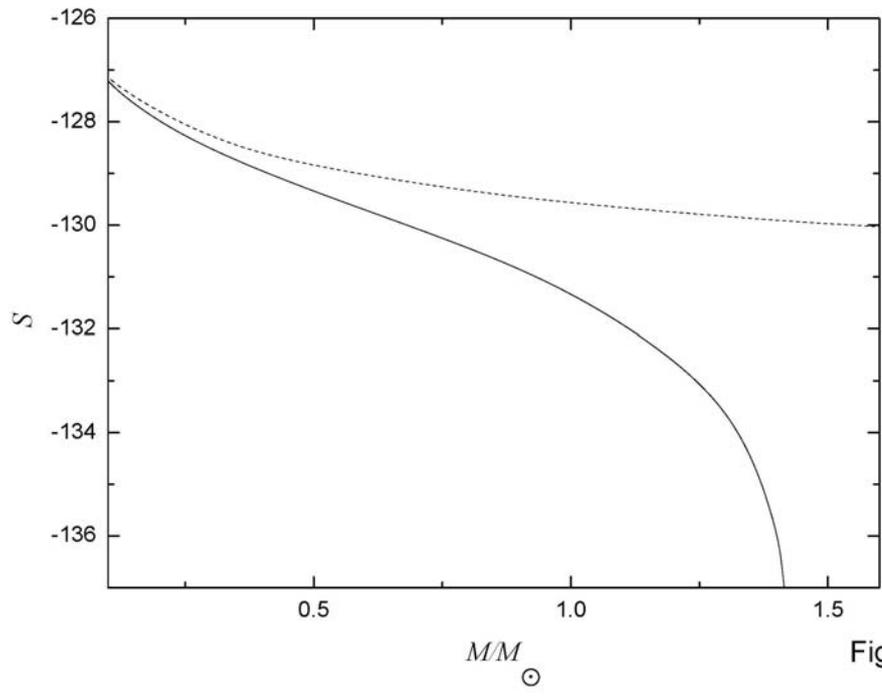

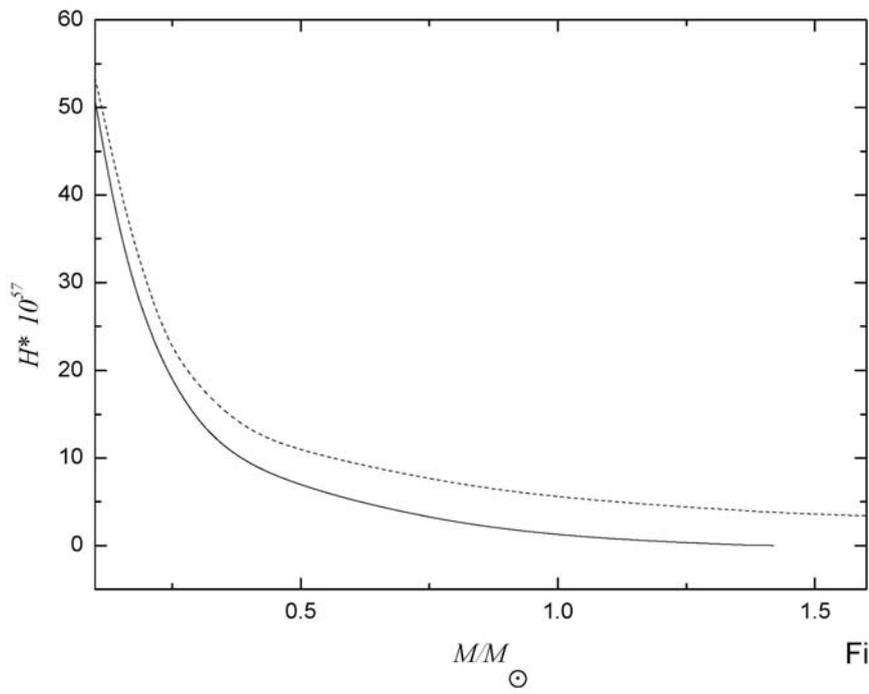

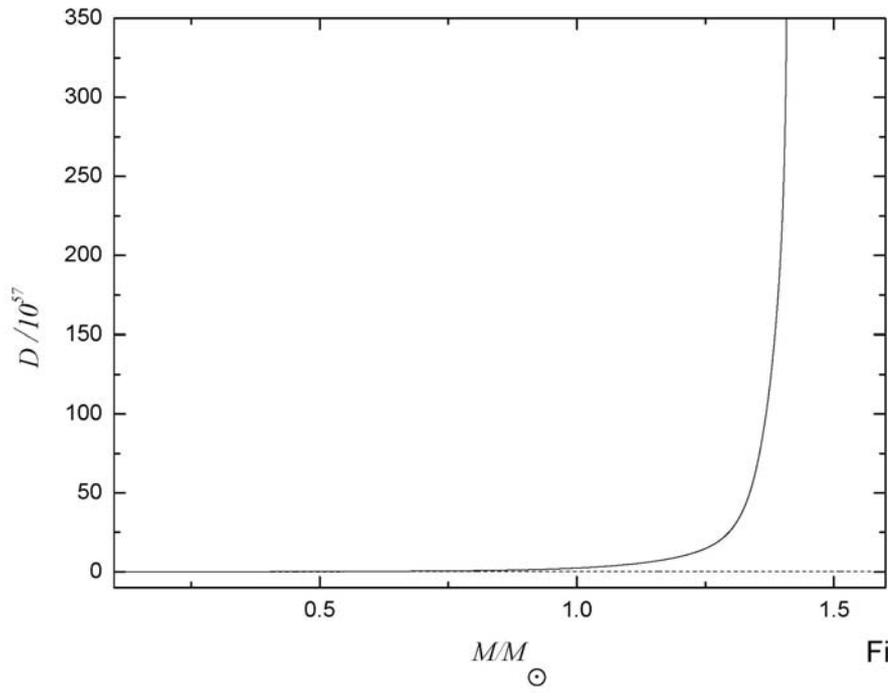

Fig.1c

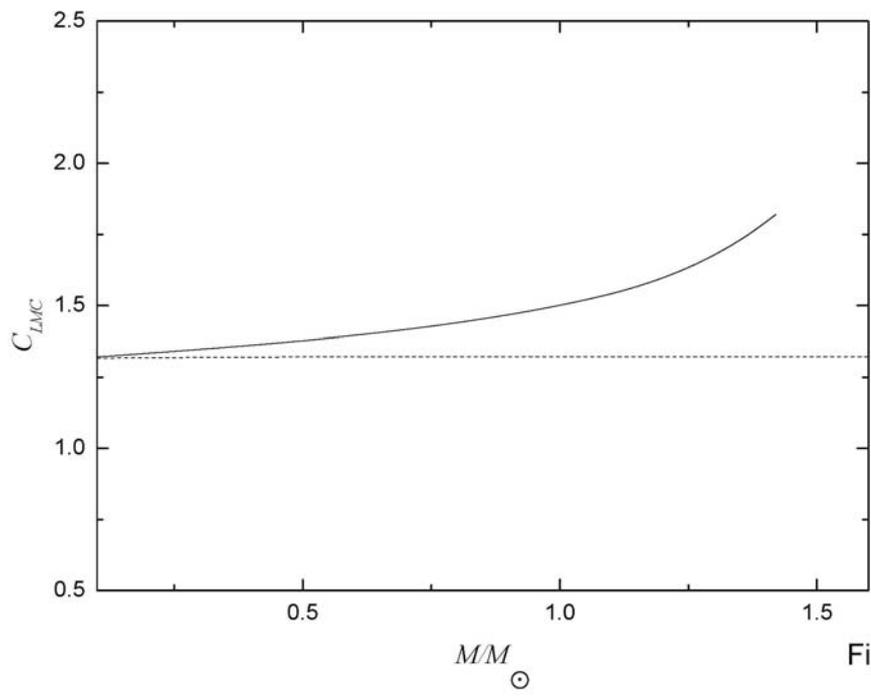

Fig.1d